\documentclass[intlimits,twoside,a4paper]{article}

\usepackage{amsmath,amssymb}
\usepackage{graphicx}
\usepackage{wrapfig}
\usepackage{siunitx}

\usepackage[T2A]{fontenc}
\usepackage[cp1251]{inputenc}

\usepackage{textcomp}

\usepackage{cmpj2}
\issue{2013}{16}{3}{33401}
\doinumber{10.5488/CMP.16.33401}

\title[Atomistic modelling of friction of Cu and Au nanoparticles adsorbed on graphene]%
{Atomistic modelling of friction of Cu and Au nanoparticles adsorbed on graphene%
}
\author[A.V. Khomenko, N.V. Prodanov, B.N.J. Persson]{A.V. Khomenko\refaddr{label1,label2}, N.V. Prodanov\refaddr{label1,label3,label4}, B.N.J. Persson\refaddr{label2}}
\addresses{
\addr{label1} Department of Complex Systems Modeling, Sumy State University, 2 Rimskii-Korsakov St., 40007 Sumy, Ukraine
\addr{label2} Peter Gr\"unberg Institut-1, Forschungszentrum-J\"ulich, 52425 J\"ulich, Germany
\addr{label3} J\"ulich Supercomputing Center, Institute for Advanced Simulation, Forschungszentrum-J\"ulich, \\ 52425 J\"ulich, Germany
\addr{label4} Department of Materials Science and Engineering, Universit\"at des Saarlandes, 66123 Saarbr\"ucken, Germany
}

\date{Received April 27, 2013, in final form May 29, 2013}
\authorcopyright{A.V. Khomenko, N.V. Prodanov, B.N.J. Persson, 2013}

\begin{document}

\maketitle

\begin{abstract}

We present classical molecular dynamics calculations of the behavior of copper and gold nanoparticles on a graphene sheet, sheared with a constant applied force $F_{\rm a}$.
The force $F_{\rm s}$ acting on the particle from the substrate depends on the material of the nanoparticles (Au or Cu), and exhibits a sawtooth dependency on time,
which we attribute to local commensurability between the metal nanoparticle surface atomic positions with the graphene lattice.
The time-averaged value of $F_{\rm s}$ (the friction force) acting on Au nanoparticles increases linearly with the contact area,
having slopes close to the experimentally observable ones. A qualitative model is proposed to explain the observed results.
\pacs 46.55.+d, 
    62.20.Qp, 
    81.40.Pq, 
    68.35.Af, 
    68.37.Ps, 
    61.72.Hh 

\end{abstract}

\section{Introduction}\label{sec1}

The atomic force microscope (AFM) and surface force apparatus have greatly expanded the possibilities to investigate tribological properties of nanoscale objects~\cite{Gnecco2007}.
In particular, the AFM has been used to study the frictional interaction between nanoparticles (NPs) on atomically flat substrates. These represent very well defined experiments
where both the size and shape of the nanoparticles and the nature of the substrate can be studied with atomic resolution both before and after sliding the particles~\cite{Ritter2005}.

A series of experiments concerned with the shear of NPs have been carried out recently~\cite{Gnecco2007,Ritter2005,Dietzel2008,Dietz2010tl,Dietz2010}.
The measurements suggest that the friction force acting on antimony NPs deposited on highly oriented pyrolytic graphite (HOPG) depends linearly
on the contact area of the NP. Another interesting result is the discovery that some nanoislands with the
contact area up to $9 \cdot 10^4$~nm$^{2}$ show very low friction forces, while other similar NPs exhibit much larger friction forces.

In spite of some theoretical efforts presented in literature~\cite{Ritter2005,Brndiar2011}, no definitive explanation of the experimental results has been proposed so far.
Classical molecular dynamics (MD) simulations of silver and nickel NPs adsorbed on graphene and sheared with constant
external force $F_{\rm a}$ were presented in reference \cite{Khomenko2010jpc}, where it was shown that the friction force acting on the nanoislands changes linearly with the contact area.
The friction force is quite sensitive to the material the NP is made of. In particular, the results suggest that Ni NPs
on graphene experience higher friction than the Ag NPs. Moreover, a sawtooth form of the substrate shear force $F_{\rm s}$ acting on the nanoparticle, versus the lateral component of the center of mass
position, was found for Ni while the corresponding curves for Ag NPs turned out to be rather irregular.
Such a behavior was attributed to the local commensurability of the counterface of Ni NPs due to the
close match of the nearest-neighbor distance in metal and the lattice constant of the graphene substrate.

In reference~\cite{Khomenko2010jpc} only two metals were considered and additional investigations are required to test how general
the previous computational results are and to prepare the foundations for a quantitative theory.
With this aim we perform classical MD simulations described in this paper. We pursue two main goals. First, we want to confirm the
results related to the effect of the material of NPs on their tribological behavior. It is clear that short-range order of atoms can play a significant role in determining
the friction force~\cite{Khomenko2010jpc}. If the distance $a$ between nearest-neighbors in metal is close to the graphene lattice
constant of 0.246~nm \cite{Castro2009}, then local commensurability of atoms is possible leading to a sawtooth type dependence of the force $F_{\rm s}$ acting on the NP from the
substrate (graphene).
The time-average  $\langle F_{\rm s} (t)\rangle = F_{\rm f}$ of the (lateral) substrate force is the friction force.
The commensurability can happen for Ni, where $a=0.249$~nm~\cite{MetalsHandbook}. On the other hand, for metals with larger values of $a$, commensurability should not occur
and this was observed for Ag with $a=0.289$~nm \cite{Khomenko2010jpc}. To confirm this hypothesis we consider here two other metals: Cu with $a=0.256$~nm and Au with $a=0.289$~nm \cite{MetalsHandbook}.

The paper is organized as follows. Section~\ref{sec2} describes the atomistic model. Results of the calculations and the discussion are given in section~\ref{sec3}. Brief conclusions are presented in section~\ref{sec4}.

\section{Model}\label{sec2}

\begin{figure}[!b]
\centerline{
\includegraphics[width=0.48\textwidth]{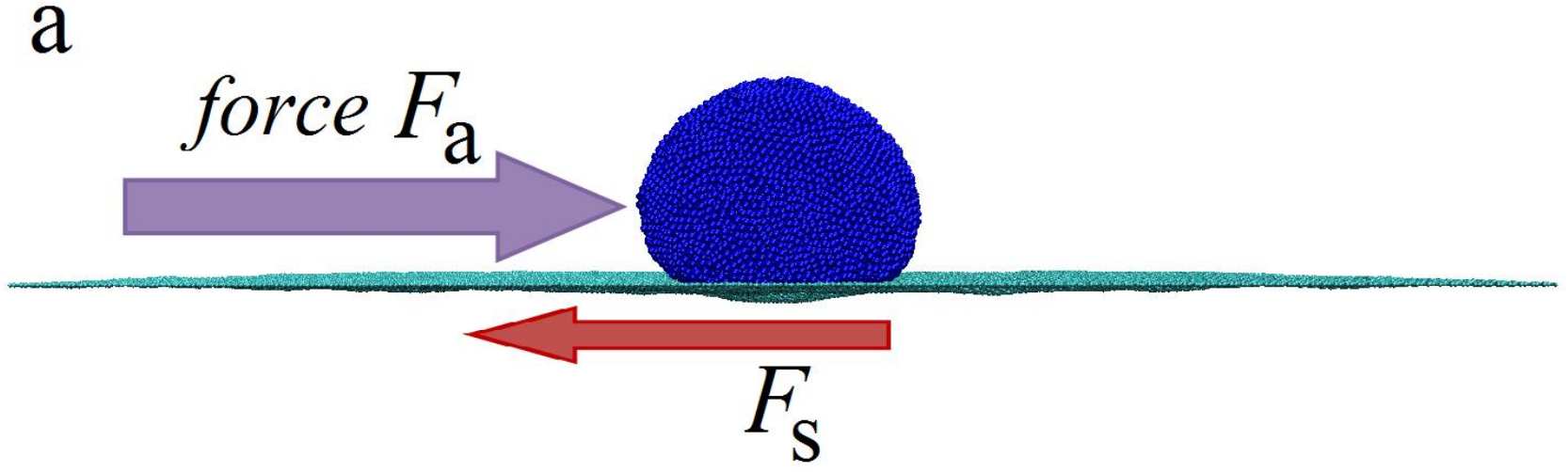}
\hspace{5mm}
\includegraphics[width=0.3\textwidth]{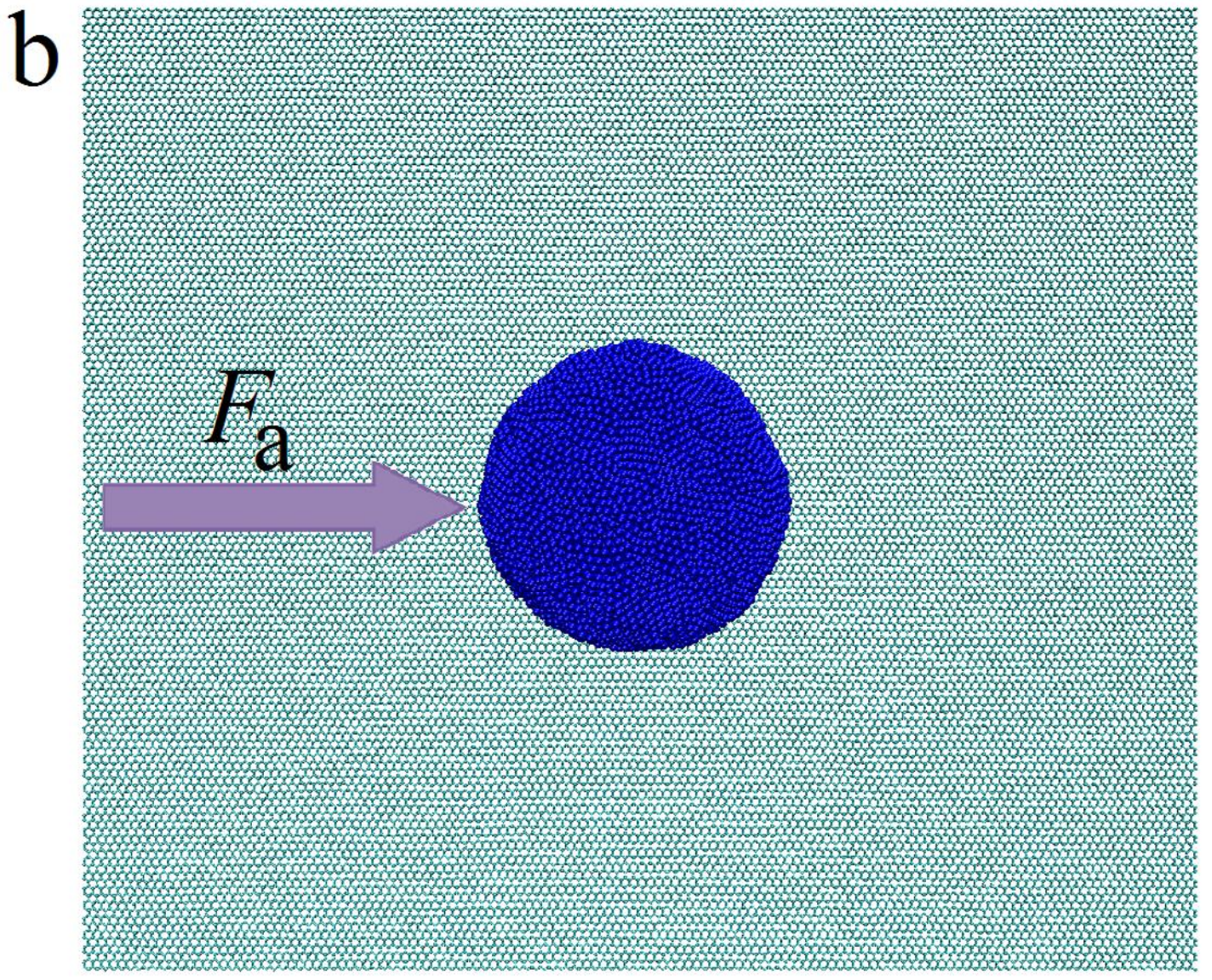}
}
\caption{(Color online) Snapshots of the formed nanoparticle containing 22000 Cu atoms: perspective (a) and top (b) views.}
\label{fig1}
\end{figure}

A detailed description of the model can be found in reference~\cite{Khomenko2010jpc}. Here, we provide only a general sketch and the details
which are specific to the current calculations. The substrate is a graphene sheet parallel to the $xy$-plane, with zigzag and armchair edges along the $x$- and $y$-directions,
respectively (see figure~\ref{fig1}; all snapshots in this work were produced with VMD software \cite{Humphrey1996}). Boundary carbon atoms along the perimeter of the graphene
layer are held rigid in order to provide a fixed position of the layer. Copper and gold nanoislands containing from 5000 up to 30000 atoms are considered.
Depending on the size of the nanoparticle, the $x\times y$ dimensions of the graphene sheet varied from about $19.68 \times 17.04$~nm to $36.40 \times 31.52$~nm.
The total number of atoms in the calculations varies from 17800 to 73808.

Interactions between carbon atoms in graphene are described by the harmonic potential~\cite{Sasaki1996}. Embedded atom method (EAM) potential is employed for forces between metal atoms~\cite{Zhou2001}. For the metal-carbon interaction the pairwise 6-12 Lennard-Jones (LJ) potential is used~\cite{Khomenko2010jpc,Sasaki1996}. The simulation code is implemented using the NVIDIA CUDA platform \cite{Meel2008,Anderson2008} which allowed us to carry out the computations on a single graphics processing unit (GPU) NVIDIA GeForce GTX 460. Algorithms for GPU based on the neighborlist technique~\cite{Griebel2007,Rapaport2004} from reference~\cite{Anderson2008} with our own algorithm for binning atoms into cells are employed. The equations of motion are integrated using the leapfrog method~\cite{Griebel2007,Rapaport2004} with a time step $\Delta t$ = 0.2~fs.


\section{Results and discussion}\label{sec3}

NPs are obtained using the procedure similar to the dewetting of thin metallic films by thermal treatments~\cite{Geissler2010} and described in reference~\cite{Khomenko2010jpc}.
Briefly, a rectangular slab of metal atoms is placed on the graphene sheet and the simulation starts.
The metal atoms assemble into a more compact structure, which is more energetically favorable than the rectangular slab.
At an appropriate time moment the system is cooled down in order to get a hemispherical shape of the NP and to prevent it from forming a ball. {During this preparation stage a single Berendsen thermostat~\cite{Griebel2007} is applied to both metal and carbon atoms.}
After the equilibrium has been reached, the external force $F_\mathrm{a}$ is applied along the $x$-direction (zigzag edge of graphene),
and the NP starts to move in this direction. This force is distributed uniformly on all metal atoms with the $x<X_{\rm CM}$, where $X_{\rm CM}$ is the
NP center of mass position along the $x$-direction. {Starting from this moment the thermostat decouples from the metal atoms and acts only on the graphene in order to diminish the effect of the temperature control on the dynamics of the system.}
We perform several series of calculations with different values of the applied force $F_{\rm a}$ acting on the NPs of the same size.
Where it is not stated explicitly, the $F_\mathrm{a}$ acting on Cu NPs consisting of 5000 and 29000 atoms is equal to 2.31~nN and 15.29~nN, respectively.
For Au NP consisting of 10000 and 25000 atoms $F_\mathrm{a}$ is 5.07~nN and 13.05~nN, respectively. We note that in our MD calculations the NPs are accelerating
since in general the applied force $F_{\rm a}$ is not equal to the substrate force~$F_{\rm s}$.

\begin{wrapfigure}{i}{0.52\textwidth}
\centerline{
\includegraphics[width=0.52\textwidth]{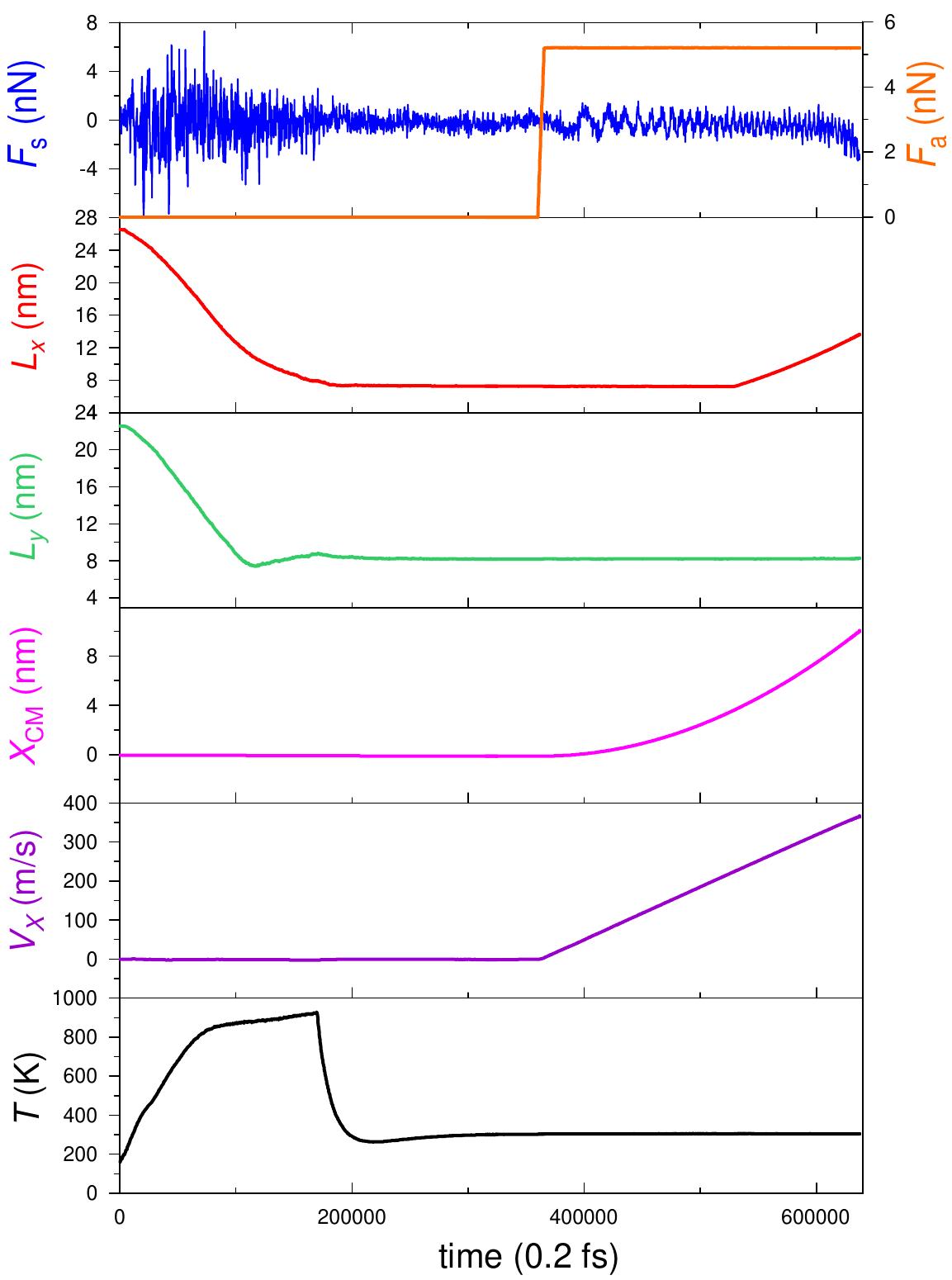}
}
\caption{(Color online) Time dependencies of the system temperature $T$, lateral position $X_\mathrm{CM}$ and velocity $V_X$ of the center of mass of the NP, total applied force $F_\mathrm{a}$,
substrate force $F_\mathrm{s}$, and lateral dimensions $L_x$ and $L_y$ obtained for Cu nanoisland containing 19000 atoms.}
\label{fig2}
\end{wrapfigure}
%
Figure~\ref{fig2} shows the time dependency of the quantities measured during the simulation of a copper NP containing 19000 atoms.
Note that once the motion of the NP is initiated, the values of $V_X$ and $X_\mathrm{CM}$ increase nearly linearly and quadratically with time, respectively.
This is possible only if the friction force acting on the NP is independent of the sliding velocity (in the studied velocity range) so that
and $F_{\rm a}-F_{\rm f}$ is time independent. However, the substrate force $F_{\rm s}(t)$ acting on the NP is not constant, but exhibits a sawtooth shape with
increasing frequency (in time) as the particle moves faster, which indicates an atomistic stick-slip motion of the NP.
We note that using MD simulations (or any other exact numerical simulation method) one cannot study the low NP sliding velocities (of order $\sim$~\SI{}{\micro\metre}/s)
prevailing in the AFM measurements, but the numerical results presented below indicate a very weak velocity dependency in the studied velocity interval.

\begin{figure}[!b]
\centerline{
\includegraphics[width=0.4\textwidth]{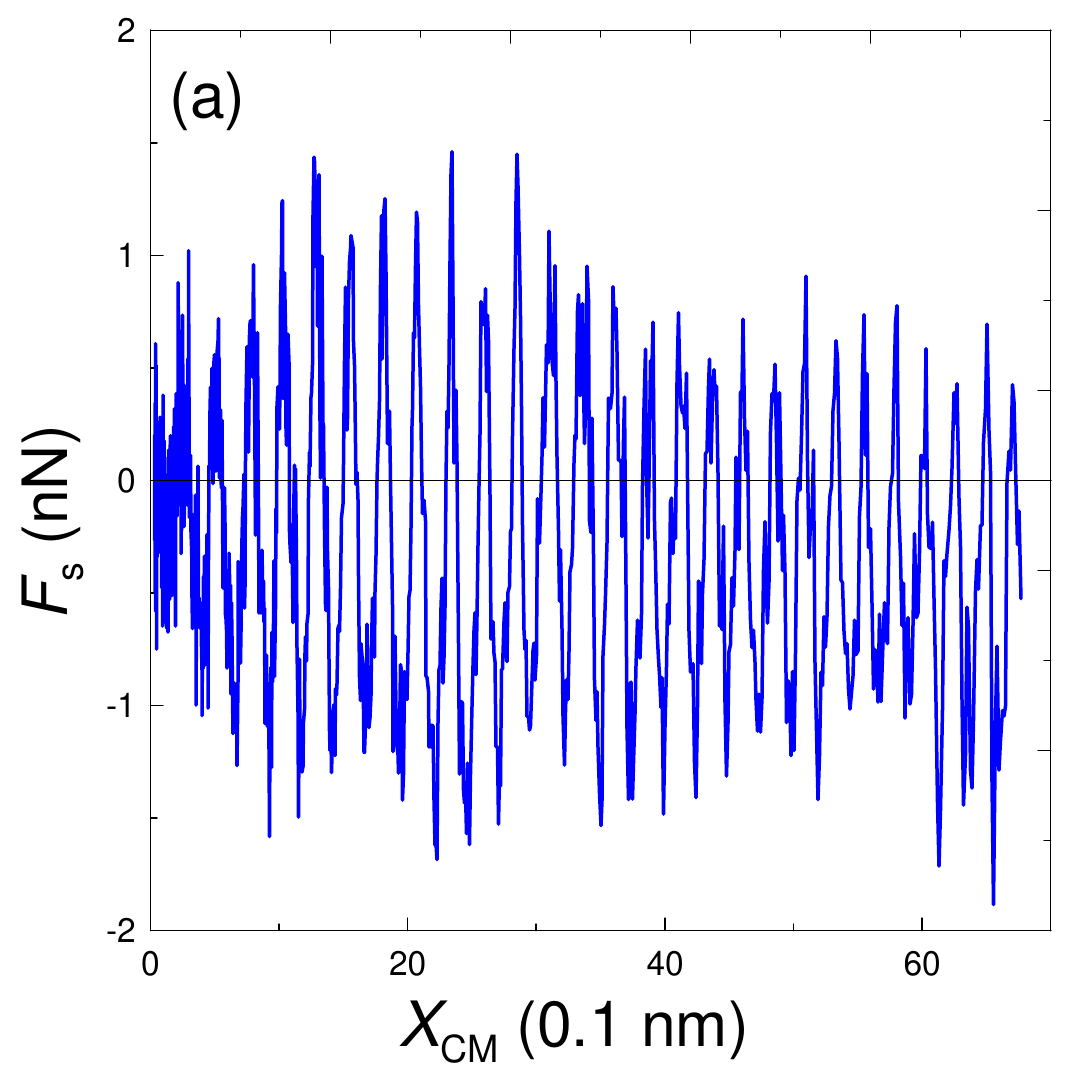}
\hspace{1mm}
\includegraphics[width=0.395\textwidth]{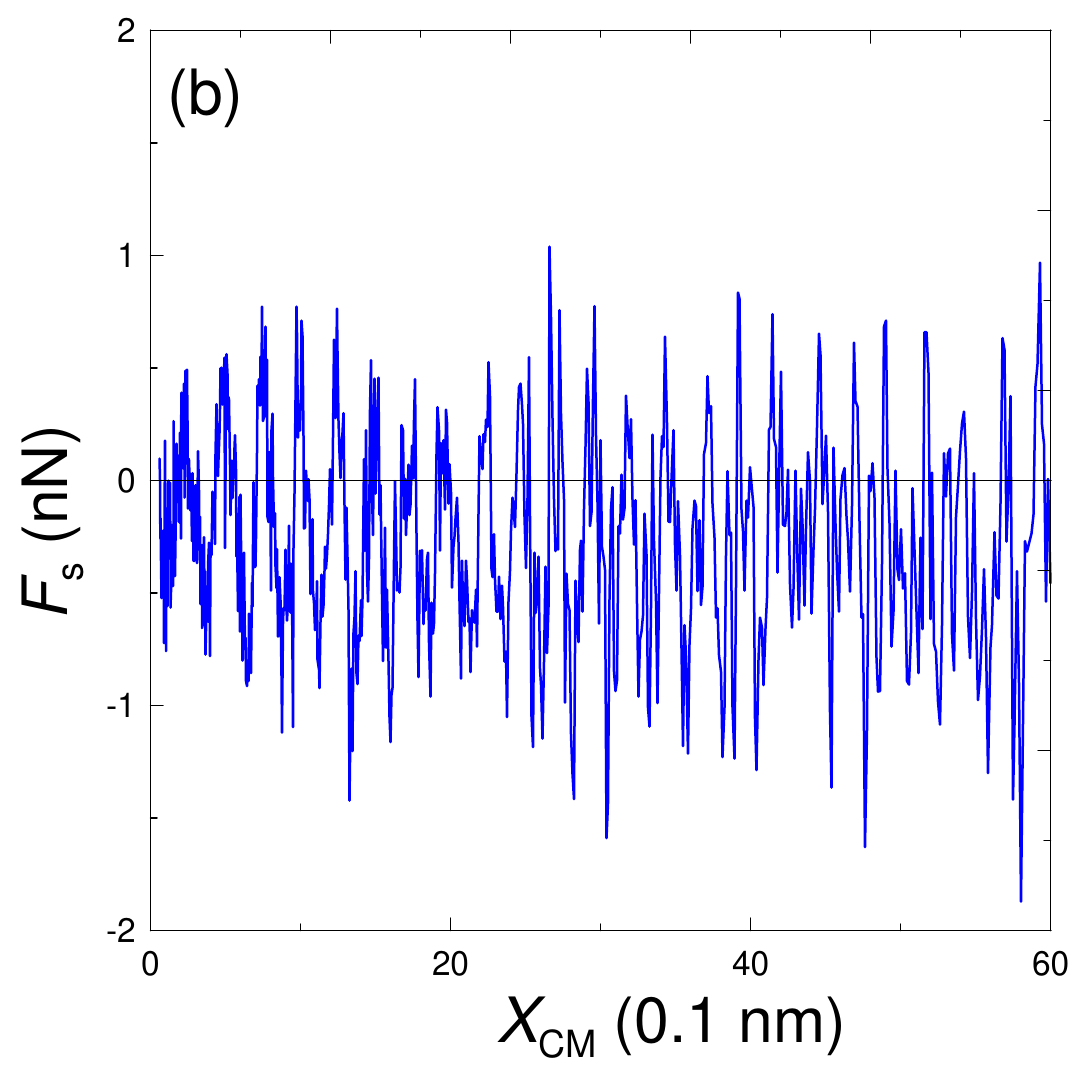}
}
\vspace{1mm}
\centerline{
\includegraphics[width=0.4\textwidth]{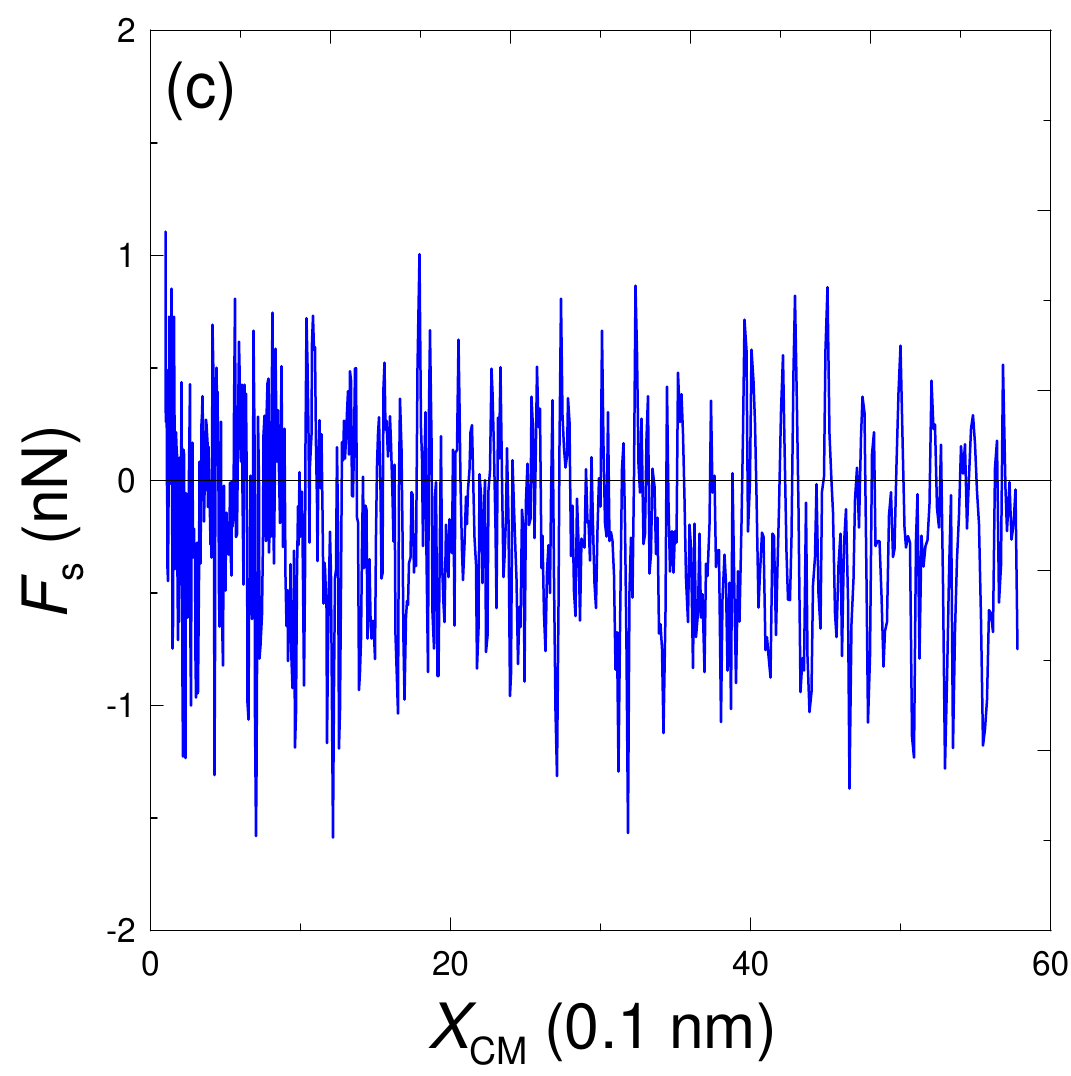}
\hspace{1mm}
\includegraphics[width=0.4\textwidth]{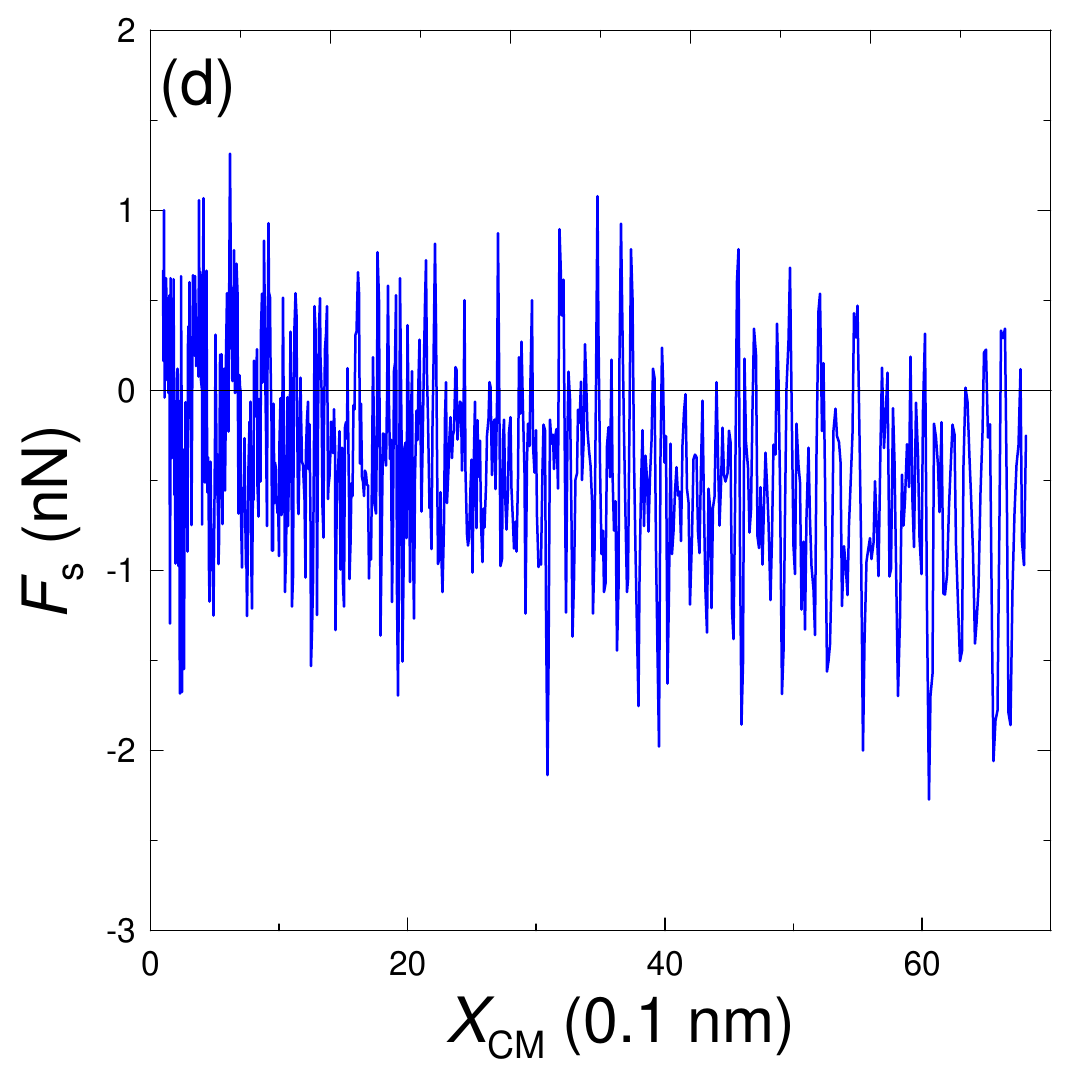}
}
\caption{(Color online) Substrate force versus the lateral position of the center of mass of the nanoparticles: Cu with 5000 (a) and 29000 atoms (b), Au with 10000 (c) and 25000 (d).}
\label{fig3}
\end{figure}

To examine the sliding dynamics in more detail, in figure~\ref{fig3} we show $F_{\rm s}$ vs $X_\mathrm{CM}$ for
several Cu and Au NPs. While for the majority of Cu NPs a sawtooth dependence of the substrate force is observed and $F_{\rm s}$ peaks are more or less regular,
the $F_{\rm s}$ vs $X_\mathrm{CM}$ curves for Au have more irregular form. The distances between the peaks for Cu are close to the value of the lattice constant of graphene.
Note that $F_{\rm s}$ takes both positive and negative values. Thus, the force acting on the NP from the graphene changes its direction during the sliding of the NP.
On average, the amplitude of the peaks for Cu is larger than for Au. The substrate force for the Cu nanoparticle exhibits fluctuations with a
periodicity close to the lattice constant as expected for a nearly commensurate contact.
For the Au nanoparticles, the substrate force fluctuates much more rapidly with the displacement coordinate $X_\mathrm{CM}$,
which is expected for a nearly incommensurate (or high-order commensurate) interface \cite{Pers_2006}.

\begin{figure}[!b]
\centerline{
\includegraphics[width=0.4\textwidth]{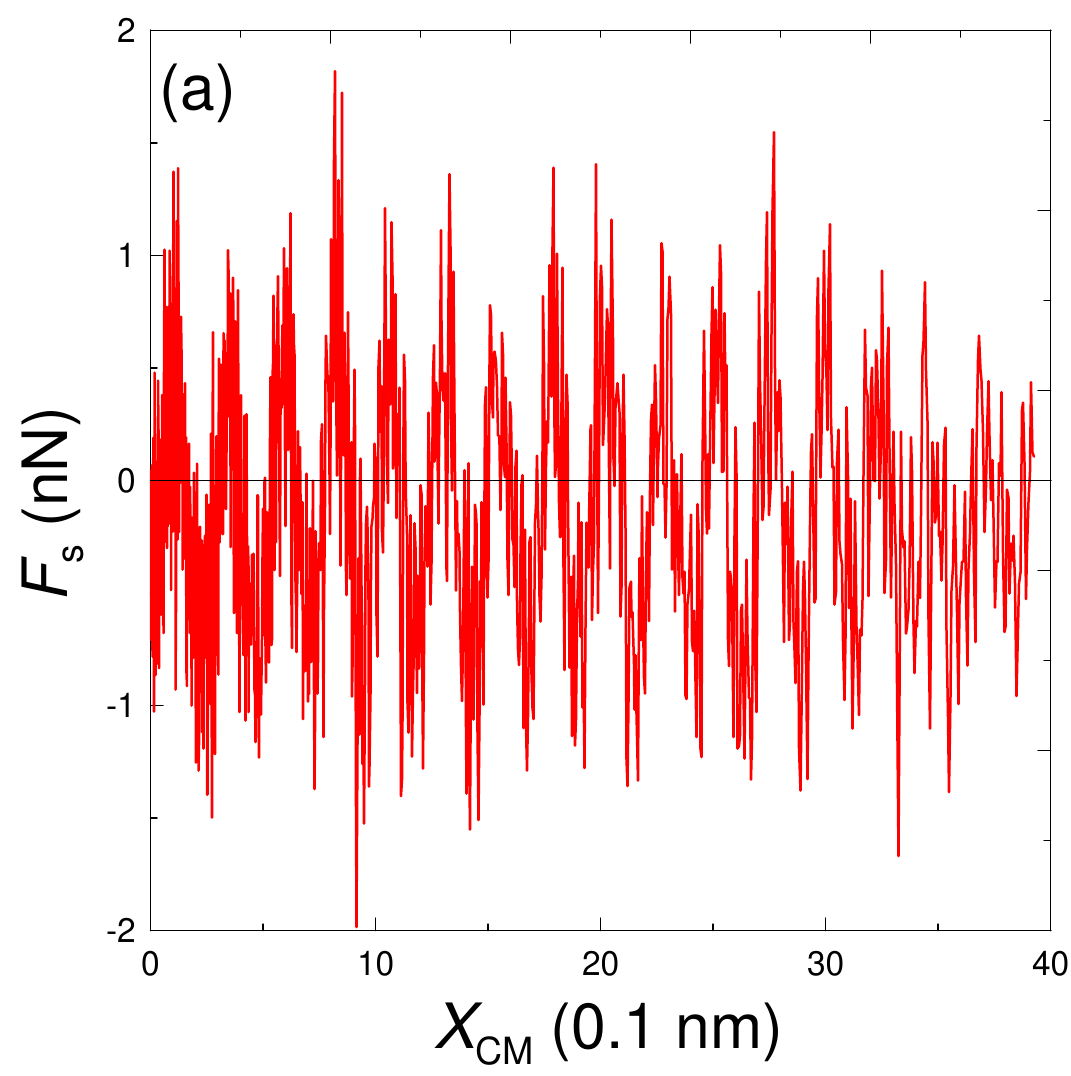}
\hspace{1mm}
\includegraphics[width=0.4\textwidth]{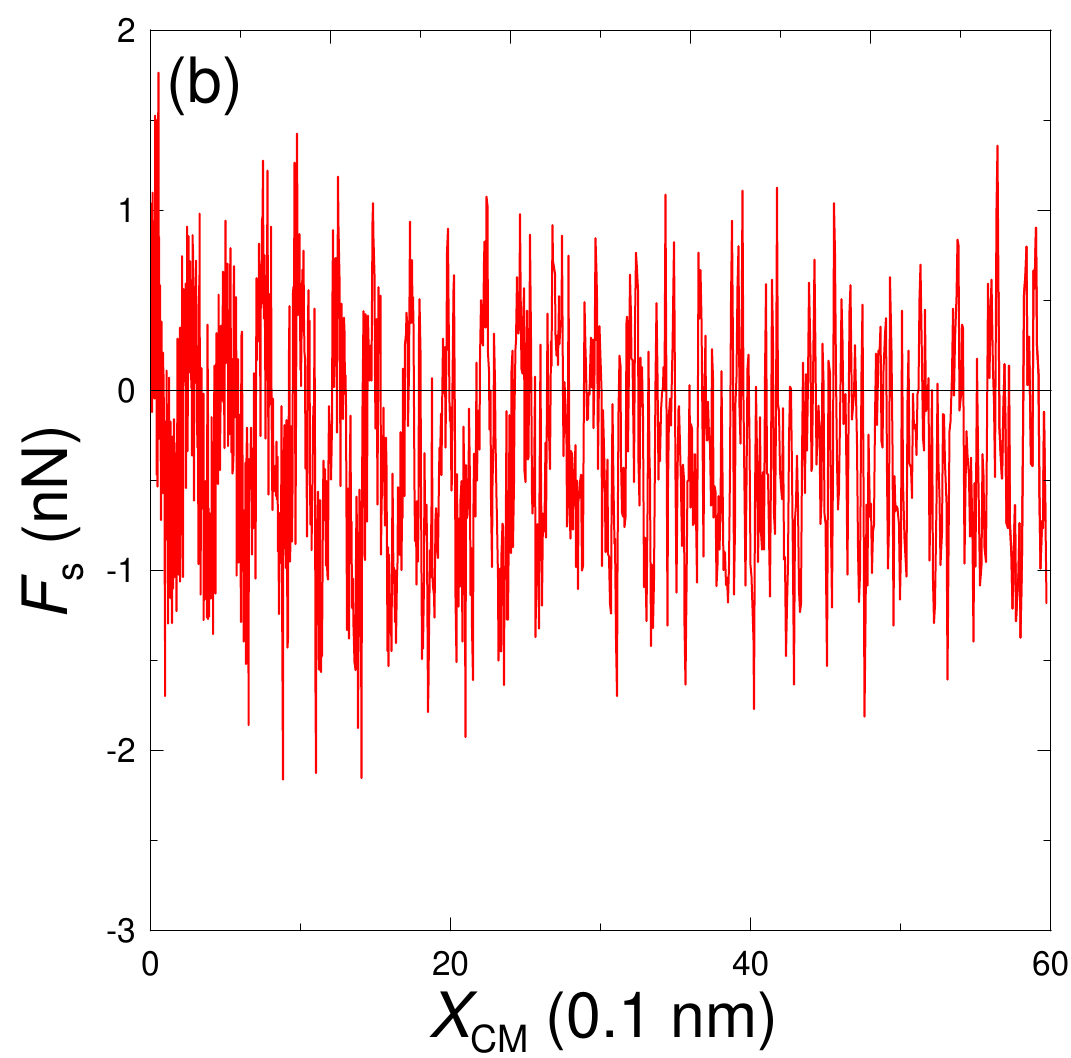}
}
\caption{(Color online) Substrate force versus the lateral position of the center of mass of the Au nanoislands at low applied force: 10000 atoms ($F_\mathrm{a}$ = 0.74~nN) (a) and 25000 atoms ($F_\mathrm{a}$ = 2.08~nN) (b).}
\label{fig4}
\end{figure}
%

Figure~\ref{fig4} presents the dependence of the substrate force $F_{\rm s}$ on the center of mass coordinate, for two different Au NPs with different applied forces $F_{\rm a}$.
Note that the curves exhibit ``fine structure'' for all sizes of the NPs, i.e., the peaks are split into high-frequency subpeaks, and that the stick-slip oscillations
are reduced or disappear for the largest $X_{\rm CM}$ corresponding to the largest sliding velocities.
To explain such a behavior one can assume that only at sufficiently low speeds the atoms on the surface of NPs can adjust to the potential energy landscape of the graphene layer.
In this way, local atomic commensurability patches can form at the interface, resulting in atomic-scale oscillations in the substrate force acting on the NP. \linebreak
By contrast, at a high sliding speed, the interfacial atoms of the NP do not have enough time to adjust to the corrugated substrate potential,
and no local commensurate regions or patches can form. This can be the reason for the irregular form of the substrate force dependencies for Au NPs at high sliding velocities.

\begin{wrapfigure}{i}{0.5\textwidth}
\centerline{
\includegraphics[width=0.4\textwidth]{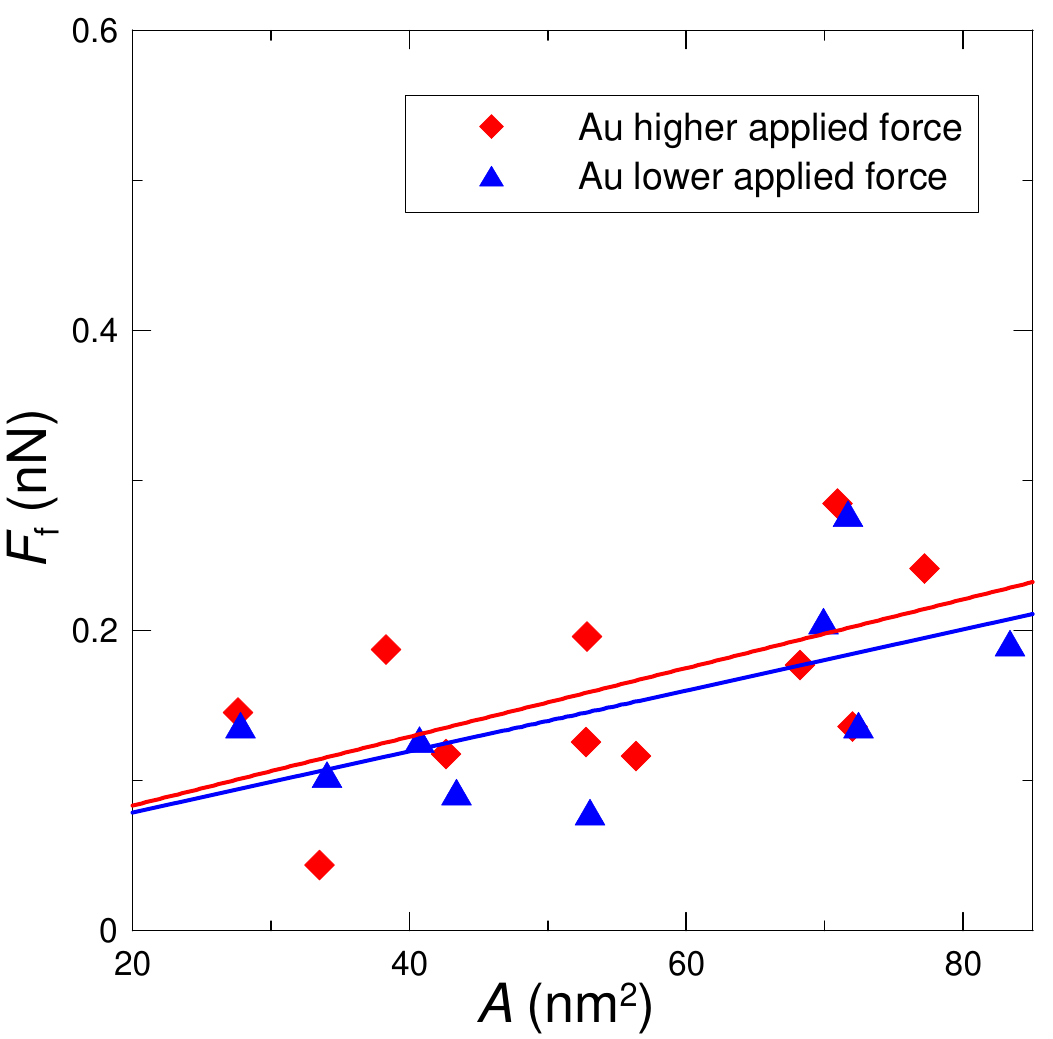}
}
\caption{(Color online) Friction force versus contact area of Au nanoparticles for different applied forces and hence different sliding velocities.}
\label{fig5}
\end{wrapfigure}
%
Figure~\ref{fig5} shows the dependence of the friction force $F_{\rm f}=\langle F_{\rm s}\rangle$ on the contact area of the NPs. {The geometry of the NPs is used in order to calculate the contact area. The shape of the NP in the $xy$-plane is approximated by an ellipse whose area is computed using the lateral dimensions of the particle.}
It is noteworthy that the change in the applied force does not significantly alter the
average value of the friction force, i.e., the friction force is nearly velocity independent in the considered velocity interval.
The slopes of the linear fits to the friction force data for Au NPs are 2.29 and 2.03~pN/nm$^2$ at a higher and lower $F_{\mathrm{a}}$, respectively.
They exceed the values of 1.21~pN/nm$^2$ obtained in MD simulations for silver~\cite{Khomenko2010jpc} and 1.04~pN/nm$^2$
from the experiments~\cite{Dietzel2008,Dietz2010}.
However, the experiments are performed at much lower sliding velocities than in the MD simulations, and the fact that the slope for the lower applied force
is slightly smaller than for the higher force is consistent with a smaller slope at the AFM sliding velocities.
Poor statistics did not allow us to reliably determine the slopes for copper NPs.
We note that if the NP atomic positions at the interface were strictly incommensurate with the
graphene lattice, one would expect a friction force roughly proportional
to the linear size of the NP contact area. The fact that the friction force is proportional to the
contact area  is consistent with a model where in the contact region the metal atoms form
commensurate patches separated by domain walls.

\begin{figure}[!b]
\centerline{
\includegraphics[width=0.4\textwidth]{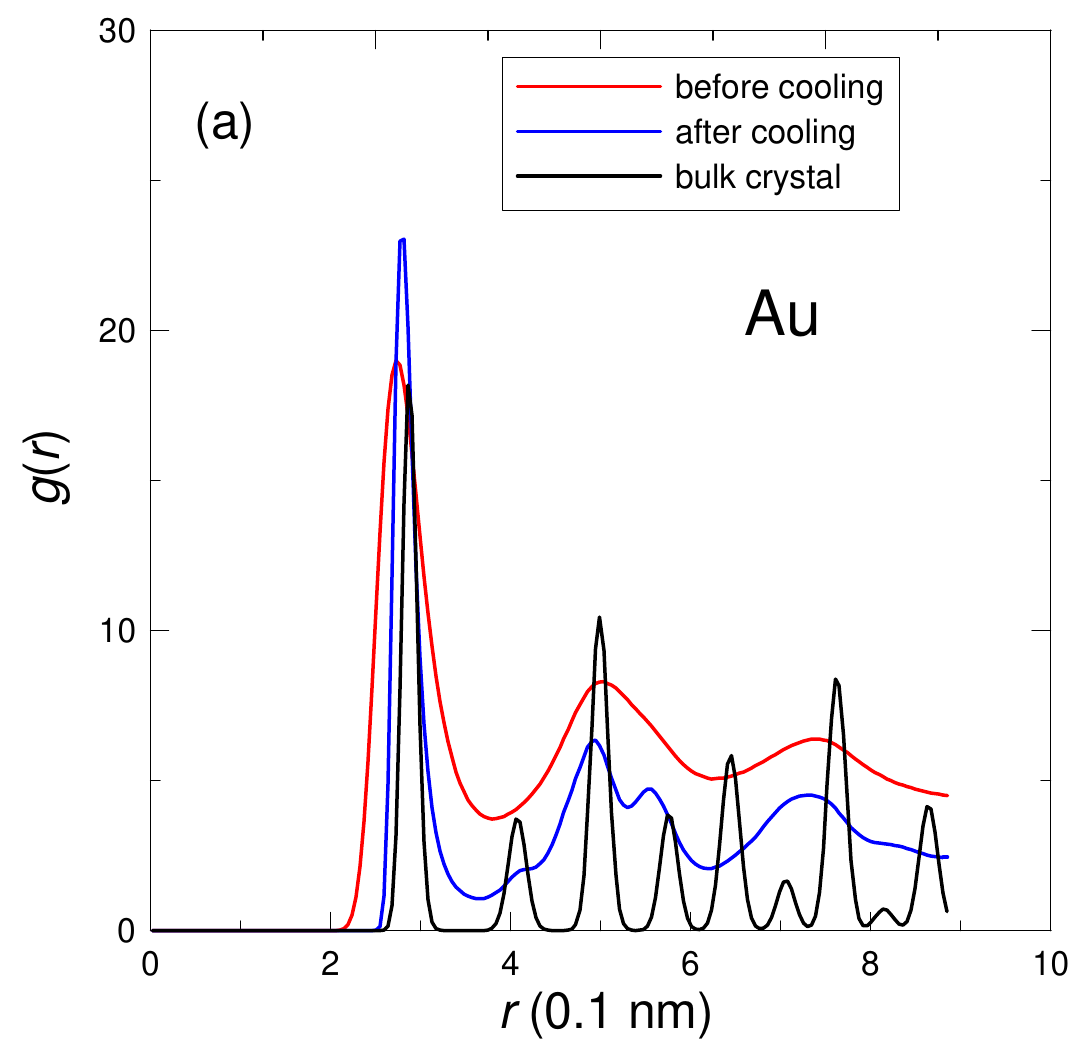}
\hspace{1mm}
\includegraphics[width=0.4\textwidth]{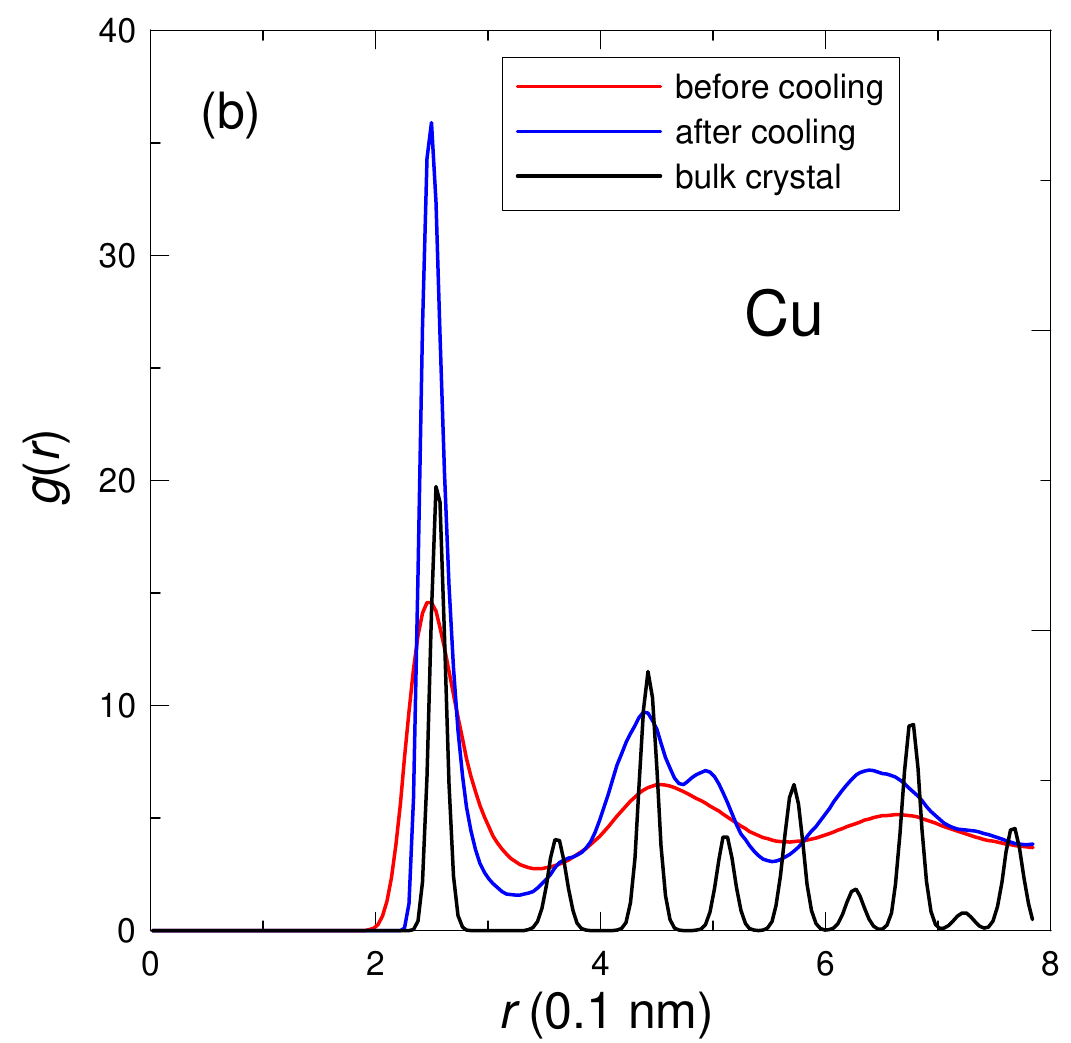}
}
\caption{(Color online) Radial distribution function obtained at different time moments for the Au (a) and Cu (b) nanoparticles containing 29000 atoms. Plots for the bulk state are obtained using the same EAM potential.}
\label{fig6}
\end{figure}

We note that thermally activated processes will become more important as the sliding
velocity decreases.
Thermal activation will tend to reduce the friction by allowing a jump over energy barriers before the driving force
has pulled the system to the top of the barriers. Thermal activation may be very important at low velocities
($\sim 1$~\SI{}{\micro\metre}/s) typically involved in AFM measurements. However, no exact atomistic calculation
(as the present MD simulations), can reach velocities as low as those typically used in most AFM measurements. Nevertheless,
most practical applications in tribology involve much higher velocities similar to those in many MD simulations.

\begin{figure}[!b]
\centerline{
\includegraphics[width=0.4\textwidth]{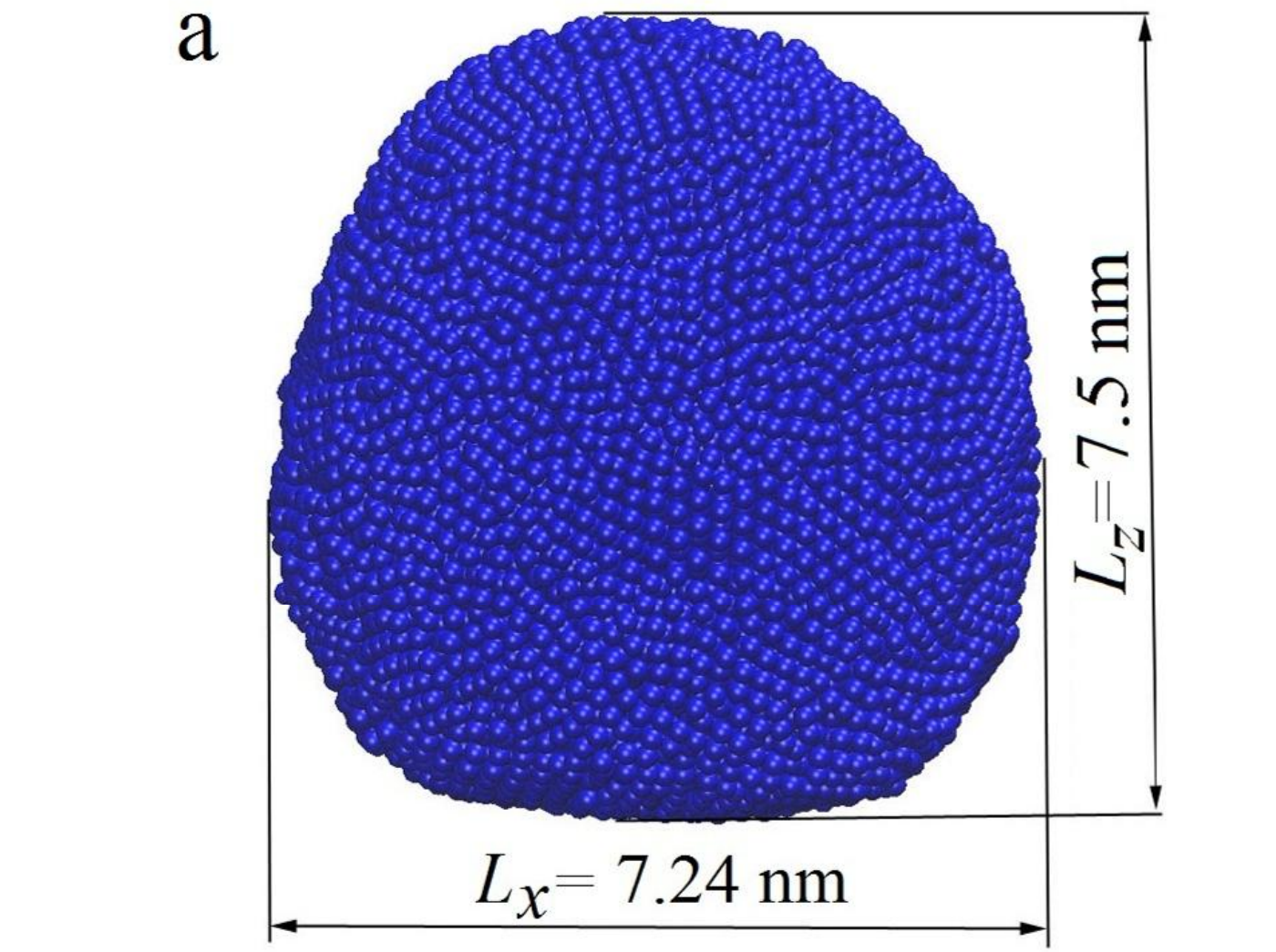}
\includegraphics[width=0.4\textwidth]{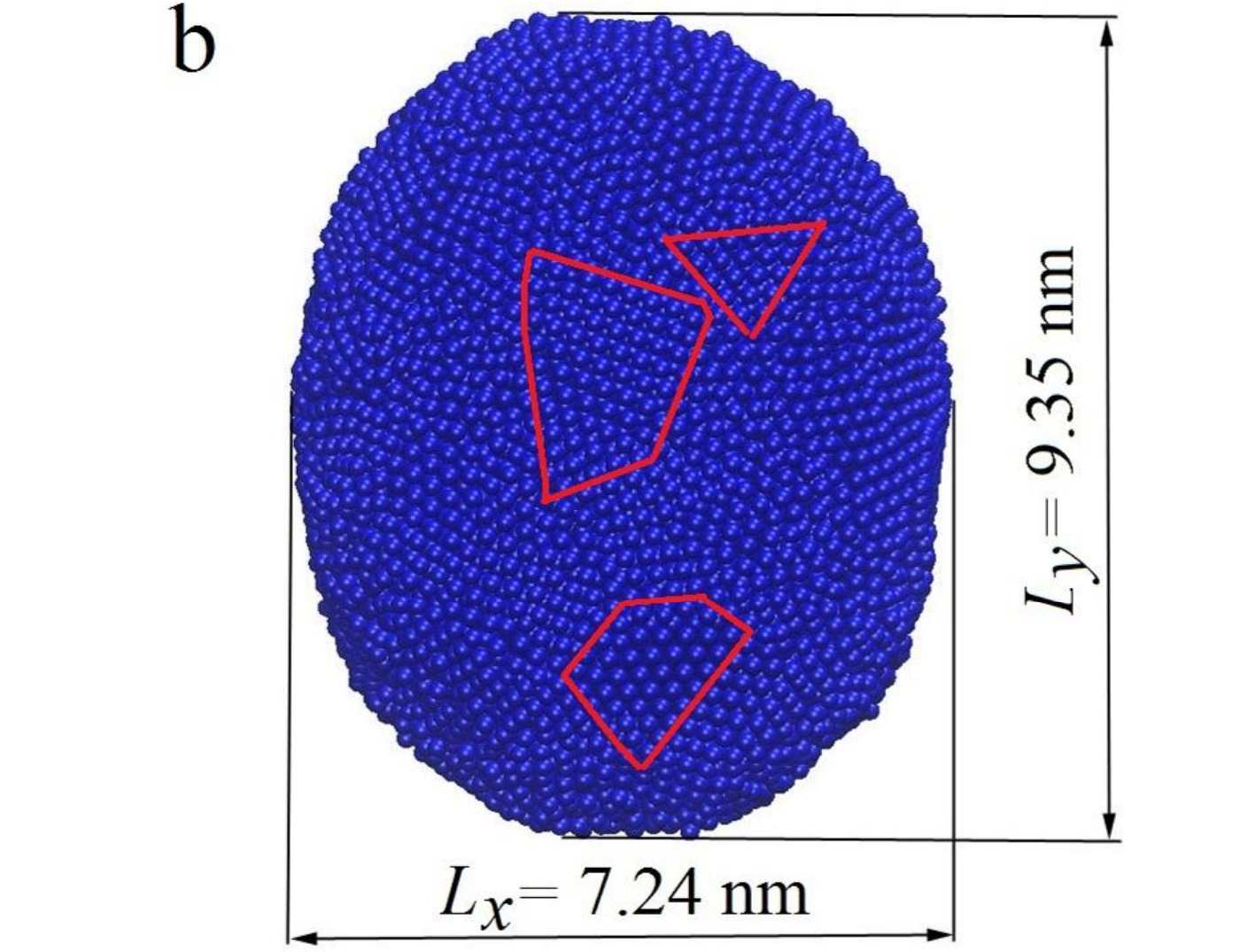}
}
\centerline{
\includegraphics[width=0.4\textwidth]{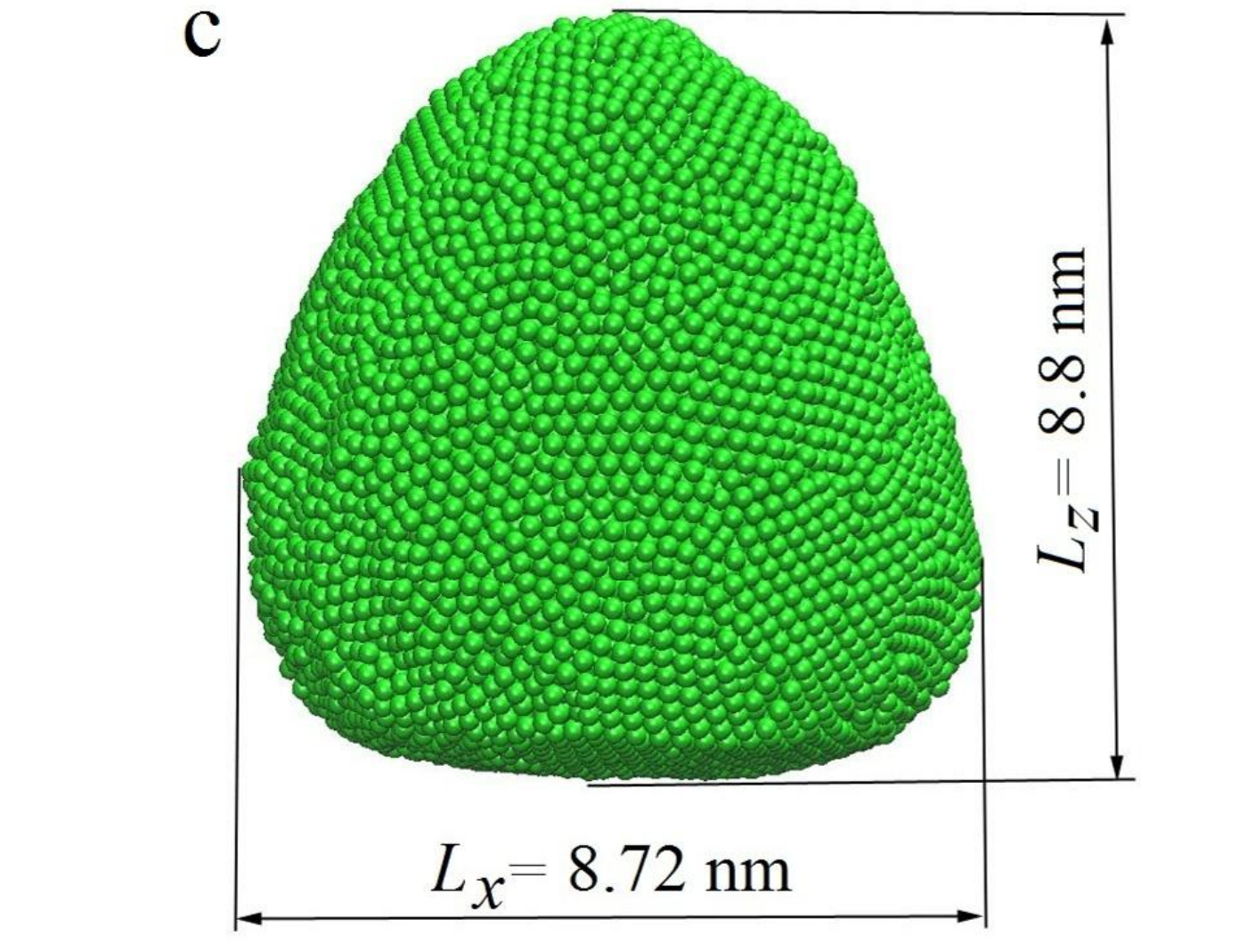}
\includegraphics[width=0.4\textwidth]{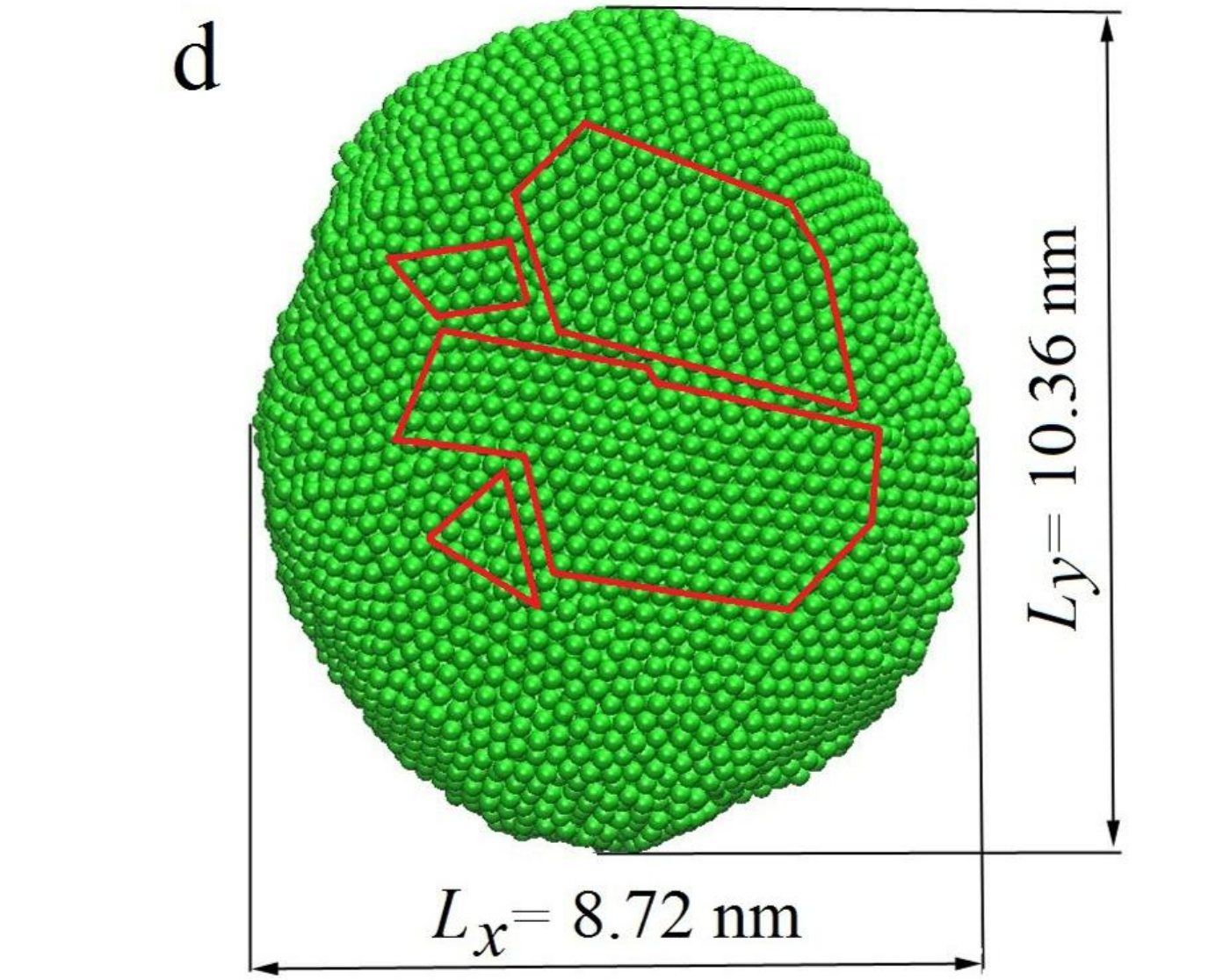}
}
\caption{(Color online) Side (a, c) and bottom (b, d) views of Cu (a, b) and Au (c, d) nanoparticles containing 25000 atoms.
Clusters of surface atoms with local order are outlined by red contours. }
\label{fig7}
\end{figure}

The structure of NPs can be analyzed using a radial distribution function (RDF) measured at different time moments, and from snapshots of the contact surface.
Typical RDFs are shown in figure~\ref{fig6}. As can be seen, RDF has a completely blurred shape after the formation of the NP for both metals,
indicating a disordered structure of the NPs. After the cooling phase some ordering occurs which
manifests in the formation of the first peak
corresponding to the distance between the nearest neighbors of about 0.256~nm and 0.289~nm for Cu and Au, respectively.
Additional smaller peaks also appear in the figure. However, they are very diffusive compared to the ideal bulk crystal.
This suggests that nanoislands have an amorphous or polycrystalline bulk structure, at least when prepared following our preparation procedure.
The structure of the NPs does not change significantly during their sliding. Visual inspection of figure~\ref{fig7} indicates that
contact surfaces of both Cu and Au NPs are not perfectly ordered but exhibit commensurate domains separated by domain walls.

The results obtained above can be summarized as follows:
\begin{itemize}
  \item a sawtooth substrate force is observed mainly for metals with the lattice constant $a$ close to the lattice constant of graphene. In our simulations it is Cu and Ni~\cite{Khomenko2010jpc};
  \item at low sliding velocities the sawtooth substrate force appears also for metals with larger discrepancy between their value of $a$ and the graphene lattice constant.
In our case this is true for Au;
  \item RDFs indicate the absence of the long-range order in the bulk for all the metals prepared using our cooling procedure;
  \item visual inspection of the surfaces (see figure~\ref{fig7}) confirms the fact that there is no long-range order of the surface atoms of the NP. However, ordered domains of atoms can be identified on the surface of NPs.
\end{itemize}

To explain the results of our computer simulations, we propose a ``patch'' model. It is based on the assumption of local commensurability of some regions of the interface.
As was mentioned above, there are domains with a local atomic order on the surface of a NP (see figure~\ref{fig7}). The domains can form commensurable regions
(``ordered patches'') at the metal-carbon interface due to some matching of metal atoms with the graphene lattice.
The commensurate regions will effectively increase the amplitude of the corrugated substrate potential experienced by the nanoparticles, and thus the friction will increase.
Such domains are formed more easily when the lattice constants of the NPs and the substrate are similar,
as well as for elastically \cite{Pers_2006} or plastically softer materials. We suggest that atomic stick-slip processes involving the  ``ordered patches'' should make the major
contribution to the net friction force of nanoislands~\cite{Khomenko2010jpc,Khomenko2008,Pogrebnjak2009}. In general, the patches are randomly distributed over the
interface and are separated by domain walls. They are quite dynamic during the motion of a NP. Every patch can be considered as a locked atom and may be the
reason for the atomic-scale oscillations in the friction force. To observe a sawtooth shape of $F_{\rm s}$ and a higher net friction force $F_{\rm f}= \langle F_{\rm s}\rangle$,
all the patches should act coherently. This situation is more probable for Cu due to the proximity of its value of $a$ to the graphene lattice constant.
Such an amplification can also occur for Au and for other metals with larger $a$, as was observed in the current work,
when the velocity of the NP is low. In this case the surface atoms have time to relax into the substrate potential energy landscape and form pinned ``patches''.
The extent to which such pinned domains can form depends not only on the lattice constants but also on the elastic or plastic properties of the solids and on the
strength of the corrugated substrate potential~\cite{Khomenko2010,Prodanov2010,Filippov2010,Gnecco2010}.

\section{Conclusions}\label{sec4}

We have performed classical MD simulations of friction of Cu and Au NPs adsorbed on graphene, and sheared with a constant external force.
The results obtained in this work confirm the earlier conclusion~\cite{Khomenko2010jpc} on a significant role of the short-range
order of atoms located on the surface of the NP. Ordered domains of atoms can be the source of the local match of the rubbing surfaces
leading to a sawtooth type dependence of the substrate force on the coordinate of the NP's center of mass.
It is shown that for Cu, especially for small NPs, sawtooth substrate force is clearly manifested.
For Au, which has larger $a$ compared to Cu, sawtooth substrate force can be observed only at a small sliding speed.
The $F_{\rm s} (t)$ exhibits ``fine structure'', i. e., larger peaks are split into a set of high-frequency peaks.

As in the previous work~\cite{Khomenko2010jpc}, the dependence of friction force on the contact area of Au NPs is linear both at higher and lower $F_{\mathrm{a}}$.
The functional form of these dependencies is not affected by changing the value of the applied force,
indicating a weak velocity dependency, which is consistent with the experimental results~\cite{Gnecco2007,Ritter2005,Dietzel2008,Dietz2010}.

A ``patch'' model is proposed in order to explain the observed behavior. The application of the model to different systems will be presented elsewhere.

\section*{Acknowledgements}

A.V.K. and N.V.P. thank the Ministry of Education and Science, Youth and Sports of Ukraine (MESYSU) for supporting
this work by the grant ``Modelling the friction of metal nanoparticles and boundary liquid films which interact with atomically flat surfaces''
(No. 0112U001380). A.V.K. acknowledges a grant from MESYSU for a research visit to the Forschungszentrum J\"ulich (Germany).


\begin{thebibliography}{99}

\bibitem{Gnecco2007}
{F}undamentals of {F}riction and {W}ear on the
  {N}anoscale, 1~Edn., Gnecco E., Meyer E. (Eds.),  {S}pringer, {B}erlin,  2007.

\bibitem{Ritter2005}
Ritter C., Heyde M., Stegemann B., Rademann K., Schwarz U.D., {P}hys. {R}ev.
  {B}, 2005, \textbf{71}, No.~8, 085405; \\ \doi{10.1103/PhysRevB.71.085405}.

\bibitem{Dietzel2008}
Dietzel D., Ritter C., M\"{o}nninghoff T., Fuchs H., Schirmeisen~A., Schwarz
  U.D., {P}hys. {Rev}. {L}ett., 2008, \textbf{101}, No.~12, 125505; \doi{10.1103/PhysRevLett.101.125505}.

\bibitem{Dietz2010tl}
Dietzel D., Feldmann M., Herding C., Schwarz U.D., Schirmeisen~A., {T}ribol.
  {L}ett., 2010, \textbf{39}, 273; \\ \doi{10.1007/s11249-010-9643-z}.

\bibitem{Dietz2010}
Dietzel D., M\"{o}nninghoff T., Herding C., Feldmann M., Fuchs H., Stegemann
  B., Ritter C., Schwarz U.D., Schirmeisen~A., {P}hys. {R}ev. {B}, 2010,
  \textbf{82}, 035401; \doi{10.1103/PhysRevB.82.035401}.

\bibitem{Brndiar2011}
Brndiar J., Turansk\'{y} R.,  Dietzel D., Schirmeisen~A., Stich I., {N}anotechnology,
  2011, \textbf{22}, 085704; \\ \doi{10.1088/0957-4484/22/8/085704}.

\bibitem{Khomenko2010jpc}
Khomenko A.V., Prodanov N.V., {J}. {P}hys. {C}hem. {C}, 2010, \textbf{114},
  19958; \doi{10.1021/jp108981e}.

\bibitem{Castro2009}
Castro-Neto A.H., Guinea F., Peres N.M.R., Novoselov K.S., Geim A.K., {R}ev.
  {M}od. {P}hys., 2009, \textbf{81}, No.~1, 109; \doi{10.1103/RevModPhys.81.109}.

\bibitem{MetalsHandbook}
{A}SM {M}etals {H}andbook, Vol.~2, {P}roperties and {S}election: {N}onferrous
  {A}lloys and {S}pecial-{P}urpose {M}aterials, 10 Edn., {A}merican {S}ociety {f}or
  {M}etals, {M}etals {P}ark, {O}hio,  1992.

\bibitem{Humphrey1996}
Humphrey W., Dalke A., Schulten K., {J}. {M}ol. {G}raphics, 1996, \textbf{14},
  No.~1, 33; \doi{10.1016/0263-7855(96)00018-5}.

\bibitem{Sasaki1996}
Sasaki N., Kobayashi K., Tsukada M., {P}hys. {R}ev. {B}, 1996, \textbf{54},
  No.~3, 2138; \doi{10.1103/PhysRevB.54.2138}.

\bibitem{Zhou2001}
Zhou X.W., Wadley H., Johnson R., Larson D., Tabat N., Cerezo A., Petford-Long
  A., Smith G., Clifton P., Martens~R., Kelly T., {A}cta {M}ater., 2001,
  \textbf{49}, No.~19, 4005; \doi{10.1016/S1359-6454(01)00287-7}.

\bibitem{Meel2008}
Van Meel J.A., Arnold A., Frenkel D., Portegies-Zwart S., Belleman R.G., {M}ol.
  {S}imul., 2008, \textbf{34}, No.~3, 259; \doi{10.1080/08927020701744295}.

\bibitem{Anderson2008}
Anderson J.A., Lorenz C.D., Travesset A., {J}. {C}omput. {P}hys., 2008,
  \textbf{227}, No.~10, 5342; \doi{10.1016/j.jcp.2008.01.047}.

\bibitem{Griebel2007}
Griebel M., Knapek S., Zumbusch G., {N}umerical {S}imulation in {M}olecular
  {D}ynamics, {S}pringer, {B}erlin, 2007.

\bibitem{Rapaport2004}
Rapaport D.C., {T}he {A}rt of {M}olecular {D}ynamics {S}imulation, 2 Edn., {C}ambridge
  {U}niversity {P}ress, {C}ambridge,  2004.

\bibitem{Geissler2010}
Geissler A., He M., Benoit J.M., Petit P., {J}. {P}hys. {C}hem. {C}., 2010,
  \textbf{114}, No.~1, 89; \doi{10.1021/jp908427r}.

\bibitem{Pers_2006}
Tartaglino U., Samoilov V.N., Persson B.N.J., J. Phys.: Condens.
  Matter, 2006, \textbf{18}, No.~17, 4143; \\ \doi{10.1088/0953-8984/18/17/004}.

\bibitem{Khomenko2008}
Khomenko A.V., Prodanov N.V., {C}ondens. {M}atter {P}hys., 2008, \textbf{11},
  No.~4, 615; \doi{10.5488/CMP.11.4.615}.

\bibitem{Pogrebnjak2009}
Pogrebnjak A.D., Shpak A.P., Azarenkov N.A., Beresnev V.M., {P}hys.-{U}sp., 2009, \textbf{52}, No.~1, 29; \\ \doi{10.3367/UFNe.0179.200901b.0035}.

\bibitem{Khomenko2010}
Khomenko A.V., Prodanov N.V., {C}arbon, 2010, \textbf{48}, No.~4, 1234; \doi{10.1016/j.carbon.2009.11.046}.

\bibitem{Prodanov2010}
Prodanov N.V., Khomenko A.V., {S}urf. {S}ci., 2010, \textbf{604}, No. 7--8,
  730; \doi{10.1016/j.susc.2010.01.024}.

\bibitem{Filippov2010}
Filippov A.E., Vanossi A., Urbakh M., {P}hys. {R}ev. {L}ett., 2010,
  \textbf{104}, 074302; \doi{10.1103/PhysRevLett.104.074302}.

\bibitem{Gnecco2010}
Gnecco E., {E}urophys. {L}ett., 2010, \textbf{91}, 66008; \doi{10.1209/0295-5075/91/66008}.

\end{thebibliography}

\ukrainianpart

\title{Атомістичне моделювання тертя Cu та Au наночастинок адсорбованих на графені}

\author{О.В. Хоменко\refaddr{label1,label2}, М.В. Проданов\refaddr{label1,label3,label4}, Б.Н.Й. Перссон\refaddr{label2}}
\addresses{
\addr{label1} Кафедра моделювання складних систем, Сумський державний університет, \\
вул. Римського-Корсакова, 2, 40007 Суми, Україна
\addr{label2} Інститут Петера Грюнберга-1, Дослідницький центр Юліха, 52425 Юліх, Німеччина
\addr{label3} Суперкомп'ютерний центр Юліха, Інститут передових обчислень,
Дослідницький центр Юліха, \\ 52425 Юліх, Німеччина
\addr{label4} Кафедра фундаментального та прикладного матеріалознавства, Університет Зарланду,  \\
66123 Зарбрюкен, Німеччина
}

\makeukrtitle

\vspace{-0.25cm}
\begin{abstract}
\tolerance=3000%
Представлено розрахунки на основі класичної молекулярної динаміки поведінки мідних і золотих наночастинок на шарі графену, що зсуваються з постійною прикладеною силою $F_{\rm a}$. Сила $F_{\rm s}$, що діє на частинку з боку підкладки, залежить від матеріалу наночастинок (Au або Cu) і демонструє  пилкоподібну залежність від часу, яка обумовлена локальною сумірністю між положеннями атомів поверхні металевих наночастинок та ґратки графену. Середнє за часом значення $F_{\rm s}$ (сила тертя), що діє на Au наночастинки, збільшується лінійно з площею контакту з нахилами близькими до експериментально спостережуваних. Запропоновано якісну модель для пояснення результатів, що спостерігаються.
\keywords нанотрибологія, молекулярна динаміка, наночастинка, сила тертя, атомно-силова мікроскопія, графен
\end{abstract}

\end{document}